\def\softd{{\leavevmode\setbox1=\hbox{d}%
\hbox to 1.05\wd1{d\kern-0.4ex{\char039}\hss}}}
\def\softt{{\leavevmode\setbox1=\hbox{t}%
\hbox to \wd1{t\kern-0.6ex{\char039}\hss}}}
\def\softl{l\kern-0.45ex\raise0.1ex\hbox{'}\kern-0.10ex}
\def\softL{L\kern-0.8ex\raise0.1ex\hbox{'}\kern0.1ex}

\documentstyle{tlp}

\def\ait{\hbox{\it A\hskip -6.6pt A}}
\def\bit{\hbox{\it I\hskip -2.5pt B}}

\title[Speedup of Logic Programs by Binarization and Partial Deduction]
        {Speedup of Logic Programs by Binarization and Partial Deduction}

\author[Jan Hr\accent23uza and Petr \v{S}t\v{e}p\'anek]
{JAN HR\accent23UZA and PETR \v{S}T\v{E}P\'ANEK \\
Department of Theoretical Computer Science and Mathematical Logic,\\ Charles University,
\\ Malostransk\'e n\'am\v{e}st\'{\i} 25,
 118 00 Praha 1, CZECH REPUBLIC\\
\email{jan.hruza@mff.cuni.cz,
petr.stepanek@mff.cuni.cz}}

\begin{document}

\maketitle

\begin{abstract}
Binary logic programs can be obtained from ordinary logic
programs by a binarizing transformation. In most cases, binary
programs obtained  this way are less efficient than the
original programs. (Demoen, 1992) showed an interesting example
of a logic program whose computational behaviour was improved
when it was transformed to a binary program and then specialized by
partial deduction.
\medskip

The class of  B-stratifiable logic programs is
defined.
 It is shown that for every B-stratifiable logic program,
binarization and subsequent partial deduction produce a binary
program which does not contain variables for continuations introduced by binarization.
Such programs usually have a better computational behaviour than the original ones. Both
binarization and partial deduction can be easily automated.
A comparison with other related approaches to program transformation is given.
\end{abstract}

\section{Introduction}

Binary programs -- programs consisting of clauses with at most one atom in the body --
 appear quite naturally when
simulating  computations of Turing machines by  logic programs.
 \cite{Tar77} introduced
the concept of binary clauses.
Since then various binarizing transformations have been defined
by \cite{M86}, \cite{SS89},
 \cite{ST89} and by \cite{TB90}.
It is not difficult to show that the last three
transformations produce programs with identical computational behaviour.
\medskip

While in the beginning, binarization was a rather
theoretical issue, later, with the advent of Prolog compilers
for programs consisting of binary clauses, it found important
applications.
\cite{T92} built a Prolog system called BinProlog  that makes use of
binarization. In a preprocessing phase, the Prolog program is binarized
(see \cite{TB90}) and the binary program is compiled using BinWAM, a
specialized
version of the Warren Abstract Machine for binary programs. BinWAM is
simpler than WAM and the size of the code of the binary program is reduced.
\smallskip

Hence, it is of practical use to investigate transformations changing a
logic program to an equivalent binary logic program.
It turned out that on some programs, binarization and partial deduction produce
programs with a better performance whereas on others, programs with a worse performance are produced.
The goal of this paper is
to describe a class of programs
 for which binarization followed by partial deduction
 produces programs with a better computational behaviour.
\bigskip

The paper is organized as follows. Section 2
presents the above mentioned transformation of logic programs to
binary logic programs. Section 3 deals with the problem of
computational efficiency of binarized programs.
In section 4, B-stratifiable programs are introduced and it is proved that the
transformation consisting of binarization and partial deduction succeeds
on these programs. This transformation usually leads to a computationally
more efficient program. Section 5 gives results and conclusions.

\medskip

We shall adopt the terminology and notation of \cite{A96}. Let $H$ be an
atom,
\medskip

\centerline{$\ait \equiv A_{1}, A_{2}, \ldots, A_{m} \mbox{ and }
\bit \equiv B_{1}, B_{2}, \ldots, B_{n},\; m, n \geq 0$}
\medskip

\noindent be (possibly empty) sequences of atoms.
We restrict
our attention to definite logic programs, that is programs consisting of
clauses $ H \leftarrow \bit $ with the atom $H$ in the head and
a sequence  $\bit$ of atoms in the body. If $\bit$ is empty, we write
simply $H \leftarrow$. A clause is called {\em binary} if it
has at most one atom in the body. A program consisting of binary clauses
is called {\em binary}.

A {\em query} is a sequence of atoms. Queries are denoted by $Q$ with possible
subscripts. The {\em empty query} is denoted by $\Box$.
A computation of a logic program starts from
a non-empty query and  generates a possibly
infinite sequence of queries by SLD-resolution steps. Maximal sequences
of queries generated by
this way are called {\em SLD-derivations}. Finite SLD-derivations are
{\em successful} if they end with the empty query, otherwise they are
{\em failed.}
\medskip

In what follows, by an {\em LD-resolvent} we mean an SLD-resolvent with
respect to the leftmost selection rule and by an {\em LD-derivation} we
mean an SLD-derivation w.r.t. the leftmost selection rule.
Similarly, an {\em LD-tree} is an SLD-tree w.r.t. the leftmost selection rule.
  By {\em continuation} we mean a (possibly empty) list of terms representing goals
\cite{SS89,T92}.

\medskip

\section{A transformation to binary logic programs}\label{B_S}

We shall describe the transformation \cite{SS89} of definite logic programs to
programs consisting of binary clauses.
We define the operator $B_{S}$ transforming the queries and the
clauses of the input program.
The resulting binary program is completed by an additional clause
$c_{S}$.
\medskip

{\bf Definition 2.1}  Given a logic program $P$, let $q$ be a new unary predicate symbol,
\bigskip

(i) for a query
\medskip

\centerline{$Q\equiv A_{1}, A_{2}, \ldots, A_{n}$}
\medskip

to $P$, let
\medskip

\centerline{$B_{S}(Q) \equiv q([A_{1}, A_{2}, \ldots, A_{n}])$}
\medskip

in particular, for the empty query, we put $B_{S}(\Box) \equiv
q([\,])$.
\medskip

(ii) for a clause
\medskip

\centerline{$C \equiv H\; \leftarrow \;  B_{1}, B_{2}, \ldots, B_{n}$}
\medskip

let
\medskip

\centerline{$B_{S}(C) \equiv q([H|Cont])\; \leftarrow \;
q([B_{1}, B_{2}, \ldots, B_{n}|Cont])$}
\bigskip
\noindent where $\mathit{Cont}$ is a {\em continuation variable}. In particular, if $C$ is a unit
clause, then $B_{S}(C) \equiv q([H|Cont]) \leftarrow q(Cont)$.

\bigskip

(iii) the clause $c_{S}$ is $q([\,])\leftarrow$
\medskip

(iv) for a program $P$, we put
\medskip

\centerline{$B_{S}(P) \equiv \{B_{S}(C)| C \in P \} \cup
\{c_{S}\}$}
\bigskip

Note that $c_{S}$ is the only unit clause of the binarized program
that provides the step $B_{S}(\Box) \Rightarrow \Box$ in successful
SLD-derivations.
\bigskip

{\bf 2.2 Example} Transformation of a program to its binary form

\begin{verbatim}
a :- b,c.              q([a|Cont]) :-  q([b,c|Cont]).
b :- d.                q([b|Cont]) :-  q([d|Cont]).
c.                     q([c|Cont]) :-  q(Cont).
d.                     q([d|Cont]) :-  q(Cont).

                       q([]).
\end{verbatim}
Note that

(i) we use a different syntax for Prolog programs and for the theory of logic programs, and

(ii) continuation variables have been introduced for binarized programs. In what follows, we will use the term {\em continuation variable } also for variables in binary programs
obtained by partial deduction of binarized programs containing clauses  such as
\begin{verbatim}
q_b([a(X,1)|Cont]) :-  q_e([b(X),c(1)|Cont]).
\end{verbatim}
\medskip
{\bf 2.3 Lemma } Let $P$ be a program and $Q$ a query. Then $B_S(P) \cup \{B_S(Q)\}$ has a successful  LD--derivation with computed answer $\theta$ iff $P \cup \{Q\}$ does.
\hfill $\Box$

This follows from the fact that for every step of any LD--derivation of $P \cup \{Q\}$,
there is a corresponding step of a corresponding LD--derivation of $B_S(P) \cup \{B_S(Q)\}$.
\medskip
\section{ Transformations and binarization}
\medskip

\subsection*{Binarization can lead to more efficient programs}

Contrary to a natural expectation --
that binarization can only slow down the computations of a program
because extra arguments and extra computation steps are involved in the
transformed program, binarization followed by partial deduction
can in some cases speed the computation of a program up
significantly.
\cite{D92} was the first to present a case study of such behaviour.
\smallskip

{\bf  Transformation steps} \hfill (1)

We consider the following steps of transformation:

1. binarization

2. partial deduction with the empty continuation [ ] (i.e, the empty list) in the top--level call (see Section 4)

3. further partial deduction with  final optimization steps such as
removing duplicate variables  \cite{ML,DS99}.

\medskip
  We will show in Example 3.1 how the
above transformation steps are applied to the
\verb|SAMELEAVES| program from \cite{D92}.
We are going to investigate
programs for which this
transformation gives more efficient programs
when applied to programs with certain syntactical features and
why it leads to programs with identical or worse performance  if applied
to other programs.
First, we shall recall the \verb|SAMELEAVES| program.
\bigskip

{\bf 3.1 Example}
The program \verb|SAMELEAVES| tests whether two binary trees have the same
sequence of leaves.
The trees with the same sequence of leaves need not be isomorphic.
\bigskip

{\bf Program} \verb|SAMELEAVES|
\begin{verbatim}
        sameleaves(leaf(L),leaf(L)).
        sameleaves(tree(T1,T2),tree(S1,S2)):-
                   getleaf(T1,T2,L,T),
                   getleaf(S1,S2,L,S),
                   sameleaves(S,T).
        getleaf(leaf(A),C,A,C).
        getleaf(tree(A,B),C,L,O):-getleaf(A,tree(B,C),L,O).
\end{verbatim}
\bigskip

As the first step of transformation,  we apply the binarizing operator
$B_{S}$ from Section \ref{B_S} and obtain the following program:

\begin{verbatim}
        q([sameleaves(leaf(L),leaf(L))|Cont]):-q(Cont).
        q([sameleaves(tree(T1,T2),tree(S1,S2))|Cont]):-
                   q([getleaf(T1,T2,L,T),
                   getleaf(S1,S2,L,S),
                   sameleaves(S,T)|Cont]).
        q([getleaf(leaf(A),C,A,C)|Cont]):-q(Cont).
        q([getleaf(tree(A,B),C,L,O)|Cont]):-
               q([getleaf(A,tree(B,C),L,O)|Cont]).
        q([]).
\end{verbatim}
\bigskip

{\bf } Then we perform the steps 2 and 3. Using
an automated partial deduction system Mixtus \cite{Sa}, we partially
deduce the binarized program with the goal

\centerline{\verb|q([sameleaves(Tree1,Tree2)])|}

\smallskip

\noindent where the continuation is empty (i.e. [ ]).

\bigskip

{\bf } Applying steps 1 and 2, we obtain the following program
\begin{verbatim}
        sameleaves1(leaf(A), leaf(A)).
        sameleaves1(tree(A,B), tree(C,D)) :-
                   getleaf1(A,B,C,D).

        getleaf1(leaf(C),D,A,B) :-
                   getleaf2(A,B,C,D).
        getleaf1(tree(A,D),E,B,C) :-
                   getleaf1(A,tree(D,E),B,C).

        getleaf2(leaf(C),A,C,B) :-
                   sameleaves1(A,B).
        getleaf2(tree(A,D),E,B,C) :-
                   getleaf2(A,tree(D,E),B,C).
\end{verbatim}

The resulting program is binary and has two specialized predicates for the two calls of \verb|getleaf|.
 Demoen showed that it is faster by approximately
40\%.
\medskip

{\bf } The \verb|SAMELEAVES| example is interesting for yet another reason.
If we skip binarization and perform only partial
deduction on the original non-binary program, we  get only an identical
copy of the logic program. On the other hand,
by binarization and partial deduction w.r.t. continuation [ ],
that is by adding no information, we get a
computationally more efficient binary program by partial  deduction.
\medskip

The program \verb|FRONTIER| below computes the frontier, i.e. list of leaves of a binary tree. It serves as an example where the
above described steps of binarization and
partial deduction do not give any significant improvement.
\smallskip

{\bf 3.2 Example } \verb|FRONTIER|
\begin{verbatim}
        frontier(leaf(X),[X]).
        frontier(tree(Left,Right),Res):-
                   frontier(Left,L1),
                   frontier(Right,R1),
                   append(L1,R1,Res).
\end{verbatim}
If we perform the above steps on this program, we do
not get a computationally more efficient program. Its performance is worse in terms of time and space.
The length of the  program obtained by binarization and partial deduction is significantly larger.
It is so due to the fact that   the partial deduction system
cannot remove calls
with a free continuation variable such as
\begin{verbatim}
  q1([append([],B,B)| Cont]):- q(Cont).
\end{verbatim}

\bigskip

{\bf 3.3 Definition} We say that the binarization and partial deduction
transformation consisting of steps 1, 2 and 3 {\em succeeds} if
it terminates and eliminates all continuation variables in steps 1 and 2.

\section{B-stratifiable programs}

  In this Section, we will define the class of B-stratifiable programs,
prove that for this class of programs the transformation consisting of steps 1 and 2 succeeds
(i.e. it eliminates continuation variables.)


\bigskip

{\bf 4.1 Definition} We say that a program $P$ is {\em B-stratifiable} if there
is a partition of the set of all predicates of $P$ into disjoint sets
\medskip

\hspace{21mm}$S_{0},S_{1}, \ldots, S_{n}$ \hfill (2)
\smallskip

\noindent called {\em strata}, such that
\medskip

(i) if there is a clause $C$ such that a predicate symbol $p$, $p \in S_{i}$
occurs in the head of $C$ and  a predicate $q,\; q \in S_{j}$ occurs in the body of the same same clause $C$, then
$i \geq j$, i.e., $q$ belongs to a lower or the same stratum,  and
\medskip

(ii) in any clause  $H \leftarrow \bit\;$ of $P$ where the predicate symbol
$p$ belongs to $  S_{i}$ from the head $H$ there is  at most one predicate symbol $q$ from the same
stratum $S_i$ in the body \bit. In this case, $q$ is the predicate
symbol of the rightmost atom in \bit.
\medskip

Then the set of strata (2) is called a {\em B--stratification} of $P$. \hfill
$\Box$
\bigskip

{\bf 4.2 Example}  Program
\bigskip

\verb|         p :- q,p.|

\verb|         q :- r,r.| \hfill    (3)

\verb|         r.|

\verb|         r :- q.| \hfill      (4)

\bigskip

is not $B$-stratifiable because  \verb|q|,\verb|r|  are mutually dependent and hence in
the same
stratum, but in the body of (3) there are two calls to  \verb|r|.
If we remove the clause (4), the program becomes $B$-stratifiable.
It suffices to take
the B--stratification $S_{1}= \{\verb|r| \}, \; S_{2}=\{ \verb|q| \}, \; S_{3}=\{ \verb|p| \}$.
It is easy to check that the program \verb|SAMELEAVES| is
$B$-stratifiable while the program \verb|FRONTIER| is not.
\bigskip

 Note that the notion of B--stratifiable programs includes several classes of programs. It
can be proved that e.g.  {\em non--recursive} and {\em binary}  programs are B--stratifiable.
As for  {\em tail--recursive} programs, in the literature we found no mathematical definition but the notions  tail--recursive and B--stratifiable are similar.
\bigskip

We will show that on $B$-stratifiable programs, the transformation consisting of
 binarization and partial deduction succeeds. $B$-stratifiable programs
can be transformed with
binarization and partial deduction into binary programs that are free of continuation variables -- and usually
more efficient. This is due to the fact that the number of terms representing goals in
continuations is bounded.
\bigskip

\subsection*{ Elimination of continuation variables}
 We shall show that for every B--stratifiable program $P$ and a query $Q$, a there is  partial deduction of  $B_{S}(P) $ w.r.t.
$B_{S}(Q)$ such that the resulting program does not contain any continuation
variables.
\bigskip

In order to do this  we shall introduce a simple partial deduction alogorithm
and prove that it terminates on B--stratifiable programs, giving a new program without continuation variables.
  Intuitively, we shall compute a partial
deduction of $B_{S}(P) $ w.r.t. a set $S$.
As the
program   $B_{S}(P)$ is binary, we can use an instance of the  general partial deduction \cite{LS91}
to remove the
continuation variables. To this purpose, it is sufficient to compute
(incomplete) LD-trees to the depth one.
(A similar technique has been used in \cite{GB90,SW95}).

To make sure that the conditions of so called $S-$closedness and independence of $S$ hold to guarantee
 termination and that the partially deduced program computes the same set of answer substitutions,
 we use the following generalization operator.

\bigskip

{\bf 4.3 Definition} We define {\em generalization operator $G$}. Let
\medskip

\hspace{7mm}$Q \equiv p_{1}(t_{1},t_{2},\ldots),
p_{2}(t_{j_{2}},t_{j_{2}+1},\ldots),\ldots p_{n}(t_{j_{n}},\ldots)$
\hfill
\medskip

\noindent be a
general (non-binary)  query and

$B_S(Q) \equiv q([p_{1}(t_{1},t_{2},\ldots),
p_{2}(t_{j_{2}},t_{j_{2}+1},\ldots),\ldots p_{n}(t_{j_{n}},\ldots)])$ the respective binarized query.

We put
\medskip

\hspace{7mm}$G(B_S(Q)) \equiv q([p_{1}(X_{1},X_{2},\ldots),
p_{2}(X_{j_{2}},X_{j_{2}+1},\ldots),\ldots p_{n}(X_{j_{n}},\ldots)])$
\bigskip
  where $X_i$ are new variables.
In particular, $G(q([\,])) \equiv q([\,])$.

Note that $B_S(Q)$ is an instance of $G(B_S(Q))$.

Furthermore, we will
 extend  $G$ so that it will be applied to sets of binarized queries and atoms. If $S$ is a set of binarized  queries, we put
$G(S)\equiv \{G(Q)| Q \in S \}$.

\bigskip

{\bf 4.4 Algorithm 1}
\medskip

{\bf Input:} Binarized program $B_{S}(P)$ and the set $\{G(B_S(Q))\}$, where $P$ is a program and $Q$ a query.

{\bf Output:} A program $New\_Prog$ with no continuation variables, a set $S$, a new query $Q'$.
\medskip

I. $S := \{\,\}$,

\hspace{4mm}$\mathit{To\_be\_evaluated} := \{G(B_S(Q))\}$,

\hspace{4mm}$Prog := \{\,\}$
\bigskip

II. {\bf While} $\mathit{To\_be\_evaluated} \neq \{\,\}$ {\bf do}
\smallskip

\hspace{7mm}a) take an atom $a \in \mathit{To\_be\_evaluated}$;

\hspace{12mm}$S := S \cup \{a\}$;
\smallskip

\hspace{7mm}b) compute partial deduction of $B_{S}(P) \cup \{a\}$
obtaining an incomplete LD-tree of depth 1 (i. e. perform one
unfolding step)
\smallskip

\hspace{12mm}$R :=$ the set of resultants.

\hspace{12mm}$B :=$ the set of bodies of resultants from $R$.
\smallskip

It follows from the fact that the program
$B_{S}(P)$ is binary, that all elements of $B$ are atoms.
\smallskip

\hspace{7mm}c) $Prog := Prog \cup R$;

\hspace{12mm}$\mathit{To\_be\_evaluated} := (\mathit{To\_be\_evaluated} \cup G(B)) - S$;
\bigskip

III. {\bf Renaming } We shall define an operator $Ren$ which
 renames each atom
\bigskip

\hspace{14mm}$q([p_{1}(t_{1},t_{2},\ldots),
p_{2}(t_{j_{2}},t_{j_{2}+1},\ldots),\ldots, p_{n}(t_{j_{n}},\ldots)])$
\medskip

in $Prog$ to
\medskip

\hspace{14mm}$q\_p_{1}\_p_{2}\_\ldots\_p_{n}(t_{1},t_{2},\ldots,
t_{j_{2}},t_{j_{2}+1},\ldots,t_{j_{n}},\ldots)$
\bigskip

obtaining the program $New\_Prog$ and  the new query $Q' :=  Ren(B_S(Q))$.
\hfill $\Box$

\bigskip

We can see that the continuation variables have been eliminated by Algorithm~1.
To show that,
we can verify that the following invariant holds during the computation of
Algorithm~1 and that Algorithm~1 will terminate.
\medskip

{\bf 4.5 Invariant} No clause in $Prog$ contains a free continuation variable.
\bigskip

Now we come to the main result of this section:
\medskip

{\bf 4.6 Theorem } Let $P$ be a $B$-stratifiable program and  $Q$ a query.
 Then

1. Algorithm~1 terminates on the input $B_{S}(P) $, $\{G(B_S(Q))\}$.

2. Let  $New\_Prog$ be the output program of Algorithm 1. Then  $New\_Prog \cup \{Ren(B_S(Q))\}$ has an LD-derivation  with a
computed answer $\theta$ iff  $P \cup \{Q\}$ does.
\smallskip

{\em Proof}   We will need a definition and two lemmas that will enable us to prove termination of the algorithm.

\medskip

{\bf 4.7 Definition } Let $P$ be a logic program, $A$ an atom and let
\medskip

\hspace{14mm}$ A_{1}, \ldots, A_{n}$
\medskip

\noindent be a sequence of atoms. Let
\medskip

\hspace{14mm}$A,\;\bit \Rightarrow A_{1}, \ldots, A_{n},\; \bit \theta$
\medskip

\noindent be an LD-resolution step of $P \cup \{A\}$ where $\bit$ denotes a (possibly empty) conjunction.
 We say that each
atom $A_{i} \in \ait,\; 1 \leq i \leq n$ is an {\em immediate successor
}
of $A$ and write $A \succ A_{i}$.
Let $\succeq$ be the reflexive and transitive
closure of the immediate successor relation $\succ$. If $A
\succeq B$, we say that $B$ is a {\em successor} of $A$.\hfill
\bigskip

{\bf 4.8 Lemma }
Let $P$ be a $B$-stratifiable logic program, let
\medskip

\hspace{14mm}$S_{0}, S_{1}, \ldots, S_{n}$ \hfill (5)
\medskip

\noindent be a B--stratification of $P$ and
let $m$ be the maximum number of atoms in the body of a clause from  $P$
and let $Q \equiv A_1, \ldots  ,A_l$ be a query and $\xi$ be an arbitrary LD-derivation of $P \cup \{Q\}$.
\medskip

\noindent Then

(i) each atom $A_i, i=1, \ldots, l$,  has at most $n*(m-1)+1$ successors
in every LD-resolvent of $\xi$.

(ii) for each query $Q'$ of at most $l$ atoms, the number of atoms in any LD--resolvent of $P \cup \{Q'\}$is at most $n*(m-1)+ l$.
Hence there is a bound on the number of atoms in LD--resolvents of  $P \cup \{Q\}$.

{\em Proof}
(i) In general, if $A$ is an atom with a
predicate symbol from a stratum $S_{k}$,

$ 1 \leq k \leq n$ then $A$ has
at most $k*(m-1)+1$ successors in every LD-resolvent in $\xi$.
Hence $n*(m-1)+1$ is a bound on the number of successors of an arbitrary atom in
every LD-resolvent in $\xi$.

(ii) follows from (i).\hfill $\Box$
\medskip

Note that  the number of elements of a continuation in any LD-resolvent of
the binarized program $ B_{S}(P) \cup \{B_{S}(Q) \}  $
is equal to the number of atoms of the corresponding LD-resolvent of
$P \cup  \{Q\} $ minus 1 because for any LD-resolvent
$A_{1},A_{2}, \ldots A_{n} $
of $P \cup \{Q\}$, the corresponding continuation in the binarized
program is $[A_{2}, \ldots, A_{n}]$. \hfill
\bigskip

{\bf  4.9 Lemma} Let $ P $ be a program and let $ Q $ be a query. Assume
that there is a bound on the number of atoms in all continuations
in computations of $B_{S}(P) \cup \{B_{S}(Q)\}$.  Then there is a bound on
the number of sequences of predicate symbols in continuations that occur
in  computations of $B_{S}(P) \cup \{B_{S}(Q)\}$, too. \hfill  $\Box$
\bigskip

{\em Proof of termination of Algorithm 1}

 Algorithm 1 terminates if the set
$\mathit{To\_be\_evaluated}$ of goals for partial deduction is empty. The
elements of this set are  atoms obtained by application of the
generalization operator $G$. To guarantee the so called $S-$closedness
condition of partial deduction (see \cite{LS91}), each goal evaluated by
partial deduction is removed from $\mathit{To\_be\_evaluated}$ and is put to $S$.
The goals from the set $G(B) - S$ are added to $\mathit{To\_be\_evaluated}$,
where  $B$ is the set of goals from the bodies of resultants obtained by
partial deduction. It follows from the definition of $G$ that it maps
any two goals with the same sequence of predicate symbols to the same
atom. It follows also that S is independent.
We assumed that $P$ is a $B-$stratifiable program, hence it
follows  from Lemma 4.8 and 4.9 that there is a bound on
the number of sequences of predicate symbols in continuations that occur
in any resultant obtained by partial deduction of $B_{S}(P) \cup \{G(Q')\}$,
where $G(Q')$ is a goal from $\mathit{To\_be\_evaluated}$. It turns out that after a
finite number of steps, $\mathit{To\_be\_evaluated}$ is empty and the computation
of the algorithm terminates.
\medskip

  {\em Equivalence of computed answer substitutions.}

First, by Lemma 2.3 we can see that $B_S(P) \cup \{B_S(Q)\}$
has a successful  LD--derivation with computed answer $\theta$ iff $P \cup \{Q\}$ does.

Second, since $Prog \cup \{B_S(Q)\}$ is  $S$--closed and $S$  is independent, it follows from Theorem  4.2 of \cite{LS91}
that the resulting program $Prog \cup \{ B_S(Q)\}$ has the same computed--answer substitutions as the binarized program $B_S(P) \cup \{ B_S(Q)\} $.

Third, it is easy to see that the same holds for the renamed

$New\_Prog \cup \{ Ren(B_S(Q))\}$. \hfill $\Box$

\medskip

Once we have obtained a binary program without the continuation variables,
 further partial deduction can be performed without a limitation on the depth of LD trees.
That partial deduction can  improve performance of the program.
 Further
improvement may be obtained by the RAF procedure \cite{DS99},
\cite{ML}.
\medskip
\subsection*{A negative result}
  Now we discuss the question  whether B--stratifiable programs are exactly those on which
this transformation succeeds, i.e. whether for every non B--stratifiable program,
the binarization and partial deduction fail to eliminate continuation variables.

  Due to the fact that Algorithm 1 abstracts w.r.t. predicate symbols only and disregards terms,
there are  non--B--stratifiable programs for which the transformation still succeeds.
For example,
if we add the following clause to the \verb|SAMELEAVES| program,
 we obtain a program which is not B-stratifiable
but  the transformation described in { Example 3.1} still may succeed when applied to it:
\begin{verbatim}
getleaf(1,2,3,4) :- getleaf(5,6,7,8), getleaf(9,10,11,12).
\end{verbatim}
  This clause which caused that the program became non--B--stratifiable
is in fact never used in LD--resolution for the query
\verb|sameleaves(X,Y)|.

  On the other hand, we will give a sufficient condition for
 programs for
which the transformation does not succeed.
 We can show that for  non--B--stratifiable programs
 for which the continuation can grow arbitrarily, Algorithm~1 does not terminate.
This class of programs is large enough to include most of reasonable non--B--stratifiable programs.
 The idea of this proposition is analogous to the idea of Theorem 4.6.
\medskip

{\bf 4.10 Proposition}
  Suppose that a program $P$ is not B--stratifiable, $C$ is a clause of $P$
  containing a recursive call not in the last position in the body,
 and let  there be an atomic  query  $Q$ such that
there is a successful LD--derivation for $P \cup \{Q\}$ in which $C$ is used at least once.

Then   Algorithm~1 does not terminate with inputs $B_S(P)$ and  $G(B_S(Q))$.

\medskip

{\em Proof}

  Let $C \equiv p(t_1,\ldots,t_n) \leftarrow
\ait,p(s_1,\ldots,s_n),\bit$ be the recursive
 clause from the assumptions of the Proposition and
 let $\ait$ and $\bit$ be  sequences of  atoms such that
 $\bit$ is not empty.
Assume that  the conditions of Proposition 4.10 hold.
We shall proceed by contradiction. Assume that  Algoritm 1
with inputs $B_S(P)$ and  $G(B_S(Q))$ terminates.

As there is  a successful LD--derivation which uses the clause $C$,
Algorithm 1 will also use the binarized clause  $B_S(C)$ and add the atom

\centerline{ $A_1 \equiv  q([\ait',p(X_1,\ldots,X_n),\bit',\ldots])$
}
 to the set
$To\_be\_evaluated.$    Note that due to the generalization
operator, $\bit$ is an instance of $\bit'$ and $\ait$ is an instance
of $\ait'$.

It follows from the assumption that Algorithm 1 with the given inputs
terminates, that it will make empty the set $To\_be\_evaluated$.

Hence,  after a finite number of steps of the Algorithm 1,
an atom

\centerline { $ A_{1}' \equiv q([p(X_1,\ldots,X_n),\bit',\ldots])$
}
  will be added to the
set $To\_be\_evaluated$.

This atom will later be selected for unfolding  and
using the clause $B_S(C)$, an atom

\centerline {$A_2 \equiv q([\ait',p(X_1,\ldots,X_n),\bit',\bit',\ldots])$
}
will be added to the set $To\_be\_evaluated$.

This process is repeated infinitely many times and Algoritm 1
does
not terminate, a contradiction. This completes the proof of
Proposition 4.10. $\hfill \Box$

{\bf 4.11 Example} For the program \verb|FRONTIER|, 
its only recursive clause and the query \verb|front(X,Y)| which meet
the assumptions of this proposition, the following atoms are added to the set $\mathit{To\_be\_evaluated}$:
\begin{verbatim}
q([front(X1,X2)]),
q([front(X1,X2),front(X3,X4),append(X5,X6,X7)]),
q([front(X1,X2),front(X3,X4),append(X5,X6,X7),front(X8,X9),
    append(X10,X11,X12)]),
q([front(X1,X2),front(X3,X4),append(X5,X6,X7),front(X8,X9),
    append(X10,X11,X12),front(X13,X14),append(X15,X16,X17)])
........
\end{verbatim}
  Hence the size of atoms that are added to the set $\mathit{To\_be\_evaluated}$ grows indefinitely and Algorithm~1 does not terminate.
\section{Results and Comparison}
We will  present results of our experiments with binarization and partial deduction and give some comparison.
It may seem that the class of B--stratifiable programs
is relatively small.
Nonetheless,
some
transformations transform programs into  B--stratifiable programs,
improve the efficiency of the program significantly
and allow for further binarization and partial deduction.
We have experimented with a set of programs taken from
\cite{L99}, \cite{ML}, \cite{D92} and \cite{A96}.\footnote{ Listings of the programs can be found at
\verb|http://kti.mff.cuni.cz/\~{}hruza/binary/|.}
We used Sicstus Prolog 3.8 running on a Linux workstation and test data of reasonable size.
The first column of the table gives the name of the program,
the second through the fourth columns  give the respective speedups induced by binarization, 
partial deduction (Algorithm 1) and final optimization. (Speedup greater than 1 means the  transformed program was faster.)

\begin{tabular}{ l c c c } \hline\hline
\em  Program &  \em binarized &
\em  algorithm 1 & \em final output \\
\hline
sameleaves  &  0.42 & 0.97 & 1.31 \\
frontier1    &  0.17 & 1.03 & 1.16 \\
permutation &  0.62 &  0.90  & 1.22\\
double-append  & 0.54 & 0.86 & 1.01 \\
applast  & 0.67 &  0.95 & 1.00\\
match-append  & 0.48 &  0.93 & 1.05\\
remove  &  0.45& 0.82 & 0.92\\
contains.lam & 0.56 &  1.02& 1.21\\
\hline\hline
\end{tabular}

We can see that binarization slows programs down (as expected.)
 Subsequent removal of continuation variables without optimization   produces programs approximately as fast as the original ones,
and the final optimization leads to speedups in some cases. On the other hand, in some cases,
left--most selection rule (fixed  by binarization) does not allow for  optimizations achievable with flexible selection rule (e.g. double-append).

{\bf Memory usage }

We include a table with memory usage data for the programs, binarized programs and  subsequently partially deduced programs. Memory usage was measured in Sictus Prolog 3.8.

\begin{tabular}{ l c c c } \hline\hline
 \em  Program  &\em  binarized &
  \em algorithm 1 & \em final program \\
\hline
sameleaves  &  1.31 & 1.22 & 0.91 \\
frontier1    &  1.34 & 1.06 & 0.95\\
permutation &   1.38& 1.16  & 0.93\\
double-append  & 1.33 & 1.18 &  0.99\\
match-append  & 1.47 & 1.24 &  0.93\\
applast  & 1.75 & 1.41 &  1.24\\
remove & 1.27 & 1.16 &  1.06\\
contains.lam & 1.30 & 1.26 &  1.20\\
\hline\hline
\end{tabular}

 This table gives relative usage of heap (global stack) space  for the  binarized programs, partially evaluated programs  and the programs transformed  in steps 1,2 and 3. We can see an increase in heap usage of approximately 35 \% and a
decrease for the transformed programs to approximately 95\%.

\subsection*{Binarization with partial deduction and other approaches}

 1) In \cite{PP97b} another  transformation  using binarization is described.
In their approach, continuations are introduced flexibly {\em during} transformation and that allows for transformation during binarization.
Their approach can be understood as complementary to ours as
it  transforms a program using unfolding, folding and
 generalization producing a binary program which can be further transformed. Our approach consits in using a straightforward
binarization which is followed by tranformation -- partial deduction.

2) Another related approach to transformation
is {\em conjunctive partial deduction} (CPD) \cite{DS99}. Unlike traditional
partial deduction which considers only atoms for partial deduction,
conjunctive partial  deduction attempts  to specialize entire conjunctions
of atoms. This approach is closely related to binarization with partial
deduction. There is a difference, however. In the present approach, a program
is first
binarized and hence does not contain any conjunctions. Then standard
partial deduction can be used. Unlike that, in conjunctive partial deduction
the conjunctions are left and the system decides on splitting
conjunctions into appropriate subconjunctions.

Another difference is given
 by the fact that once a program is binarized, the selection rule is fixed and
no reordering of atoms can take place after binarization.

On some programs, conjunctive
partial deduction profits from reordering of atoms during partial deduction
as it treats clause bodies as conjuctions and not as sequences of atoms.

  It is a natural question whether on B--stratifiable programs, binarization followed by partial deduction always gives similar results as conjunctive partial deduction.
We shall show that it  is not the case.
We shall give an example of a program on which transformation consisting of binarization and  partial deduction cannot  give as good a result as conjunctive partial deduction.

\medskip

{\bf 5.1 Example} Program \verb|DOUBLEAPPEND|
\begin{verbatim}
double_append(X,Y,Z,W) :-
       append(X,Y,V),
       append(V,Z,W).

append([X|Xs],Y,[X|Zs]) :- append(Xs,Y,Zs).
append([],Y,Y).
\end{verbatim}
For this  program, CPD
 uses flexible selection rule in the construction of SLD--trees, eliminating
a second traversal of the first list. Such an optimization is not achievable
 using the standard leftmost selection rule in binarization with partial deduction but is possible in conjunctive partial deduction.

Conjunctive partial deduction approach may be somewhat more difficult to
control
but it gives greater flexibility and applicability.

It follows from our experiments with the ECCE conjunctive partial deduction system \cite{DS99}
 that  binarization with partial deduction and conjunctive
partial deduction (CPD) yields similar results when applied to the B-stratifiable
programs where optimization is achievable without atom reordering.
 A summary of results can be found in the next table:

\begin{tabular}{ l c  } \hline\hline
   Program  &  Time of ECCE/bin + PD
   \\
\hline
sameleaves  & 1.00 \\
frontier1    &  1.15\\
permutation & 0.84  \\
match-append & 0.59  \\
double-append & 0.61  \\
\hline \hline
\end{tabular}
The first column of the table gives the name of program, the second one gives ratio of run--time
for the output of conjunctive partial deduction system ECCE and binarization with partial deduction (greater than 1 means that binarization with partial deduction was faster.)
 The output programs are similar yet not always identical (for  the \verb|SAMELEAVES| profram ECCE produced the same progragram as binarization with partial deduction.)
For most programs, ECCE produced faster output. In some cases this is owed to its flexible computation rule (e.g. double-append.)

In general, conjunctive partial deduction can do all that binarization and partial deduction can.

While classical partial deduction cannot handle conjunctions (as needed e.g. for the \verb|SAMELEAVES| program),
 binarization followed by partial deduction cannot use flexible selection rule (as in the  \verb|DOUBLEAPPEND| program)
 or split a conjunction in more parts,
conjunctive partial deduction is the strongest of these transformation techniques.
\bigskip

{\bf Acknowledgements} We would like to thank Jan Hric and anonymous referees for their suggestions and comments.

\end{document}